\documentstyle[epsfig,aps,twocolumn]{revtex}

\newcommand{\rv}{{\bf r}}

\newcommand{\beq}{\begin{equation}}
\newcommand{\eeq}{\end{equation}}
\newcommand{\bea}{\begin{eqnarray}}
\newcommand{\eea}{\end{eqnarray}}
\newcommand{\up}{\uparrow}
\newcommand{\down}{\downarrow}

\renewcommand{\(}{\left(}
\renewcommand{\)}{\right)}

\newcommand{\commentout}[1]{{}}
\begin{document}
\draft
\preprint{}
\title{Particle number fractionization of an atomic Fermi-Dirac gas in an
optical lattice}
\author{Janne Ruostekoski$^{1}$, Gerald V. Dunne$^{2}$ and Juha
Javanainen$^{2}$}
\address{$^1$ Department of Physical Sciences, University of
Hertfordshire, Hatfield, Herts, AL10 9AB, UK\\ $^2$ Department of
Physics, University of Connecticut, Storrs, CT 06269, USA }
\date{\today}

\maketitle

\begin{abstract}
We show that a dilute 2-species gas of Fermi-Dirac alkali-metal
atoms in a periodic optical lattice may exhibit fractionization of
particle number when the two components are coupled via a coherent
electromagnetic field with a topologically nontrivial phase profile.
This results in fractional eigenvalues of the spin
operator with vanishing fluctuations. The fractional part can be
accurately controlled by modifying the effective detuning of the
electromagnetic field.
\end{abstract}


Particle number fractionization is a remarkable phenomenon
in both relativistic quantum field theory and condensed matter systems
\cite{jackiw,anderson}. Jackiw  and Rebbi \cite{JAC76,NIE86} showed
that for a fermionic field coupled to a bosonic field with a
topologically nontrivial soliton profile, the fermion number can be
fractional. The noninteger particle number eigenvalues may be
understood in terms of the deformations of the Dirac sea (or the hole
sea) due to its interaction with the topologically nontrivial
environment. In this paper we propose a manifestation of this phenomenon
in the atomic regime, using an optically trapped Fermi-Dirac (FD) atomic gas
\cite{OLS98}.
Fractional fermion number has been demonstrated previously in the condensed matter regime in 1D conjugated
polymers by Su, Schrieffer and Heeger \cite{SU79,HEE88,NIE86}.
Fractionally charged excitations are also fundamental to the fractional quantum Hall effect \cite{laughlin}, but the fractionization mechanism is very different from that in the polymers and in the atomic gas in this
paper. Our dilute atomic gas has a possible
advantage, compared to condensed matter systems, in the sense that the
interatomic interactions are weak and well-understood, and there now
exists a wide range of atomic physics technology to detect, manipulate,
and control atoms by means of electromagnetic (em) fields.

We study a two-species atomic FD gas
in a 1D optical lattice, coupled to an em field
with topological
properties similar to a soliton, or a phase kink. We show that this
topologically nontrivial coherent background field results in
fractionization of the particle number operator eigenvalues for the
fermionic atoms. Also, the spin operator has fractional eigenvalues
with vanishing fluctuations.

The background field is generated by means of a coherent
em field inducing transitions between the two fermionic components occupying
different internal levels, e.g., by using experimentally realized
technology of rapidly rotating laser beams \cite{AND01,MAT99}.
In the low energy limit we demonstrate the
one-to-one correspondence of the Hamiltonian for the FD atoms to the
Jackiw-Rebbi relativistic Dirac Hamiltonian describing fractionization
in quantum field theory. This is related to fractionization in the
polyacetylene polymer systems, where the linearized lattice vibrations are
coupled to the electron dynamics which becomes analogous to that for a
relativistic Dirac equation exhibiting fractional particle number.
We also show how our proposed system could be generalized to higher spatial dimensions
to represent the fractionization in relativistic 2+1 and 3+1
D quantum field theories.

Tremendous progress in experiments on cold
trapped alkali-metal atomic gases has allowed numerous studies with
Bose-Einstein condensates and the cooling of a FD gas to the quantum
degenerate regime \cite{DEM99}. Recently, condensates have been loaded to
a periodic optical lattice \cite{AND98}, a system which is expected to
exhibit interesting dynamical phenomena \cite{JAC98}. Similar
experimental progress is anticipated for FD gases.

The vacuum state in an
atomic condensate can support topological excitations
including defects, such as solitons \cite{AND01}
and vortices \cite{MAT99}, as well as SU(2) Skyrmion textures 
\cite{RUO01}. In this paper we show that in atomic gases also
the vacuum state {\it itself} may display nontrivial topological quantum
numbers with a strong analogy to the vacua encountered in relativistic
quantum field theories.

The phenomenon of fractional fermion number is best illustrated by
the following 1+1 dimensional Dirac hamiltonian \cite{JAC76,NIE86} for a
2-component spinor $\Psi(x)$ coupled to a bosonic condensate $\varphi(x)$,
which can be taken to be a static classical background field:
\beq H=\int dx\, [c\hbar\Psi^\dagger\sigma^2{d\Psi\over idx}+\hbar
g\varphi\Psi^\dagger\sigma^1\Psi+mc^2
\Psi^\dagger\sigma^3\Psi]\,.
\label{dir}
\eeq
Here $\sigma^i$ denote the Pauli spin matrices, $g$ the coupling
coefficient, $m$ the fermionic mass, and $c$ the velocity of light. We
show below that the atomic FD hamiltonian can be written in this form.

We assume the bosonic field has a doubly-degenerate ground state with
constant field $\varphi(x)=\pm\gamma$. The vacuum then exhibits a
spontaneously broken reflection symmetry
$\varphi\leftrightarrow -\varphi$. For $m=0$, the Dirac Hamiltonian
has a charge conjugation symmetry, so that for every eigenvalue
$\epsilon$ there exists an eigenvalue $-\epsilon$, and the corresponding
eigenfunctions are paired according
to $\Psi_{-\epsilon}=\sigma^3\Psi_\epsilon^*$. The fermion particle number
operator is
\beq N\equiv {1\over2}\int dx\, [\Psi^\dagger(x),\Psi(x)]\,.
\eeq
In the free vacuum, the fermion particle number vanishes.

The soliton background $\varphi(x)$  interpolates between the two vacua:
$\varphi(\infty)=-\varphi(-\infty)=\gamma$. For
$m=0$, charge conjugation symmetry is preserved, so that positive and
negative energy states are paired, but in addition to continuum modes
there is now also a zero-energy bound state localized at the soliton jump.
This state is charge self-conjugate and results in a doubly-degenerate
soliton sector vacuum. The number operator in the presence of the soliton
reads \cite{NIE86}:
\beq
N=a^\dagger a-1/2+\int dk \,(b^\dagger_k b_k-c^\dagger_k c_k)\,,
\eeq
where $b_k$ and $c_k$ denote annihilation operators for continuum fermion
and antifermion modes (respectively), while the operators $a$ and
$a^\dagger$ couple the two degenerate zero-energy ground states. The
ground-state soliton states possess fractional fermion numbers $\pm1/2$.
The Hamiltonian is diagonal in the number representation and {\it all} the
fermion eigenstates display half integral eigenvalues with vanishing
fluctuations. The fractional part of the fermion number has a topological
character: it is insensitive to local deformations of the bosonic
field, depending only on its asymptotic behavior. For $m\neq0$, charge
conjugation symmetry of the Dirac Hamiltonian (\ref{dir}) is broken and
the positive and negative energy states are no longer coupled in a simple
way. The particle number of the soliton vacuum is $\langle
0|N|0\rangle =-1/\pi
\arctan{(\hbar g\gamma/m c^2)}$ and may exhibit arbitrary fractional
eigenvalues \cite{GW}.

In our scheme to realize particle number fractionization in atomic
gases we consider neutral FD atoms loaded in a periodic optical lattice.
The confining optical potential is induced by means of the ac Stark
effect of the off-resonant laser beams \cite{AND98}. We assume a FD gas
with two internal levels $\up$ and $\down$ coupled via an
em-induced transition. The coupling could be a far-detuned optical
Raman transition via an intermediate atomic level, a microwave, or a rf
transition.
Furthermore, we assume that the two species experience optical
potentials which are shifted relative to each other by $\lambda/4$, where
$\lambda$ is the wavelength of light of the confining optical lattice.
This is realized, e.g., when the laser beam is blue-detuned from the
internal transition of the atoms in level $\up$, and red-detuned by the
same amount from the internal transition of the atoms in level $\down$. A
simple example 1D lattice potential in that case is $V_\up (x) =
V_0\sin^2(k x)$, and $V_\down (x) = -V_0\sin^2(k x)$. The neighboring
lattice sites represent atoms in different internal levels and are
separated by a distance $\lambda/4$. The
Hamiltonian for this two-species FD gas is
\bea
H/\hbar &=& {\delta\over 2} \sum_i \( \alpha_i^\dagger \alpha_i -
\beta_i^\dagger \beta_i \) -\sum_{k\, {\rm odd}} \( \kappa
\alpha_k^\dagger
\beta_{k+1} + {\rm H.c.}\) \nonumber\\ && -\sum_{l\, {\rm even}} \(
\kappa \alpha_{l+1}^\dagger \beta_{l} + {\rm H.c.} \) \,.\label{ham}
\eea
Here $\alpha_i$ and $\beta_j$ denote annihilation operators for
atoms in levels $\up$ and $\down$, at lattice sites $i$ and $j$,
respectively. The em-induced coupling between the two internal states is
described by $\kappa=\int d^3 r \psi^*_\up(\rv-\rv_k)
\Omega(\rv)\psi_\down(\rv-\rv_{k\pm 1})$, and $\delta$ stands for the
effective detuning between the levels. The em-coupled terms are the
analogues of the hopping terms in the corresponding polymer hamiltonian
\cite{NIE86}. The mode functions of the individual lattice sites (Wannier
functions) are denoted by
$\psi_j(\rv-\rv_i)$. We assume that the em coupling between the
internal levels with frequency $\Omega(\rv)$ is the only transition
mechanism for the atoms between neighboring lattice sites and
therefore we ignored the direct tunneling. For simplicity, we also
ignore the $s$-wave scattering between the two FD species.

To produce fermion number fractionization, we propose
to take the coupling frequency $\Omega(\rv)$ to be a phase-coherent
superposition of a standing em field along the $x$ axis and a field
with the spatial profile $\varphi(x)$:
\beq
\Omega(\rv)= i{\cal V}(\rv)\,[\sin(2kx)+\varphi(x)]\,, \label{ome}
\eeq
with $k\equiv 2\pi/\lambda$ \cite{opt}. We show  that such a
coupling frequency converts the atomic lattice Hamiltonian (\ref{ham})
into the Dirac Hamiltonian (\ref{dir}) in the continuum limit. This is
similar to the 1D polymer case \cite{HEE88,NIE86}, but the
physics is very different: in our atomic system the kink $\varphi(x)$
appears in the em-induced coupling between the internal atomic states,
and not as a physical domain wall kink.

\begin{figure}
\begin{minipage}{3.1cm}\vspace{-3mm}
\epsfig{width=3.1truecm, file=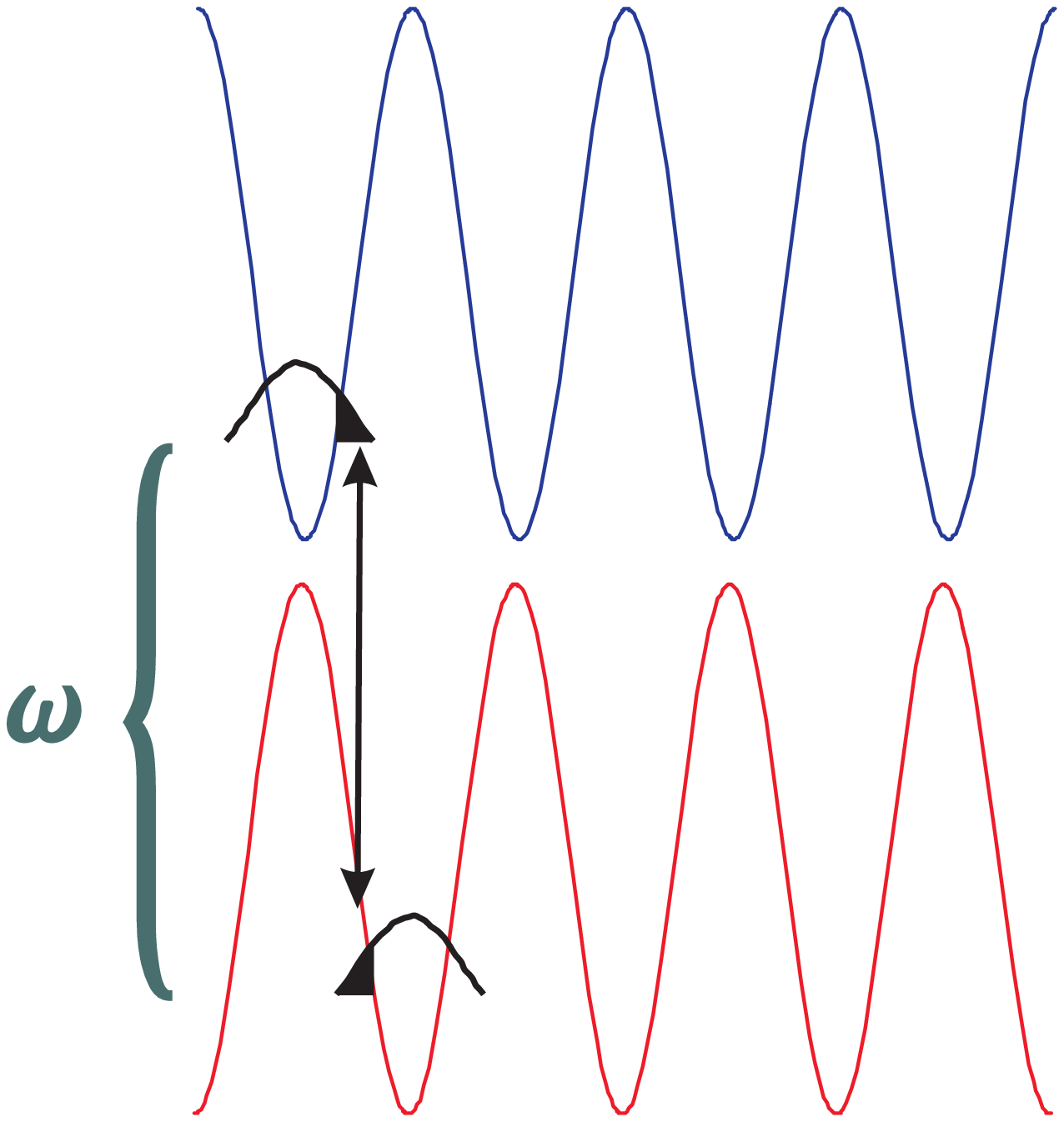}
\end{minipage}
\begin{minipage}{3.8cm}
\epsfig{width=3.8truecm,file=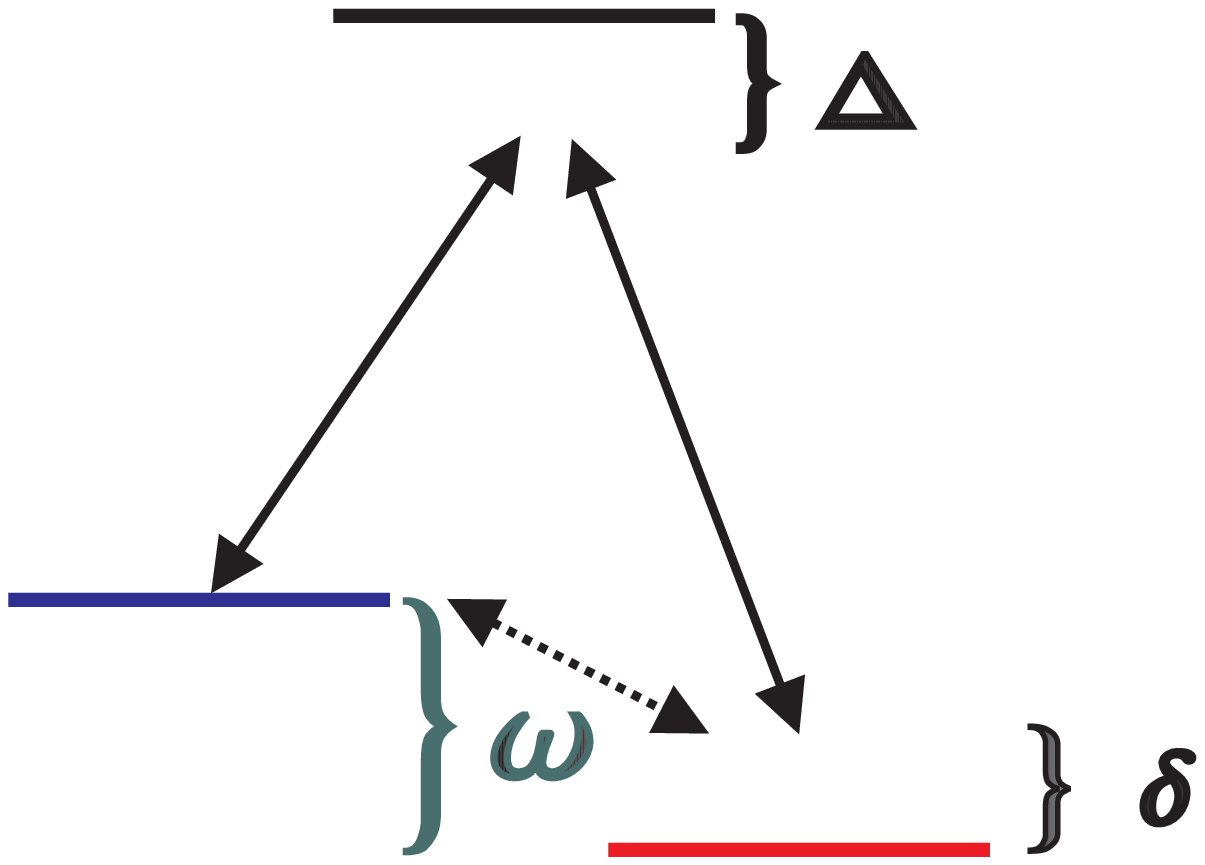}
\end{minipage}
\caption{The energy diagram of the two-species Fermi-Dirac gas in an
optical lattice. The atoms occupy two different internal levels and
experience different periodic optical potentials shifted by $\lambda/4$
(on the left). The coupling between the neighboring lattice sites,
representing different spin components, is induced by two-photon Raman
transitions, or by a superposition of the Raman transition and a one-photon
microwave or rf transition (dotted line). The transition region is denoted
by the dark shaded area representing the overlap region of the atomic wave
functions. The energy difference between the two components in
the overlap region is denoted by $\omega$. The two-photon transition is
far-detuned by $\Delta$ from an intermediate atomic level (on the right).
  }\vspace{-2mm} \label{f1}
\end{figure}

 For simplicity, we assume that
$\eta(x)\equiv \int d^3 r\,\psi^*_\up(\rv-\rv_k){\cal V}(r) \psi_\down
(\rv-\rv_{k\pm 1})$ does not change its sign over the length of the
lattice.  The purpose of the standing wave in Eq.~{(\ref{ome})} is to
introduce an  alternating sign between the neighboring lattice sites.
Such a coupling may be prepared by a two-photon optical Raman transition
(Fig.~\ref{f1}). The strength of an off-resonant two-photon Rabi
frequency in the limit of large detuning, $\Delta$, from the intermediate
state is $\Omega\propto{\cal R}_1 {\cal R}_2/\Delta$, where ${\cal R}_i$
denote the Rabi frequencies in the individual transitions
\cite{JAV95}. For two standing-wave one-photon couplings displaced from
one another by $\lambda/4$, the two-photon Rabi frequency is
$\Omega\propto\sin(kx)\cos(kx)\propto\sin(2kx)$.

We choose the field profile $\varphi(x)$ to exhibit a phase jump
of $\pi$ at $x=0$. The phase jump represents a topological phase
singularity, or a phase kink. This type of a coupling might be
produced either by making use of the (rf or microwave) transition
between the spin states, or by means of an optical Raman
transition other than the one used to produce the $\sin (2kx)$
standing wave. A phase profile with topological properties similar
to $\varphi(x)$ could be prepared, e.g., by means of a standing
microwave $\propto\sin(q x)$ with $q \ll k$. It should also be
possible to shape the wave fronts of the coupling lasers to
produce a two-photon transition with a desired phase jump. To
avoid rapid phase variation at the length scale $\lambda$, one
could use either laser beams co-propagating along the $x$ axis, or
beams with the wave vectors nearly perpendicular to $x$.
Alternatively, the coupling could possibly also be obtained by a
dc magnetic field $B(x)\propto x$ as explained in
Ref.~\cite{PU01}. An em field, topologically similar to
$\varphi(x)$, was also used to create solitons in atomic
condensates \cite{AND01}. Unlike in Ref.~\cite{AND01} we could
assume that the coupling field {\it itself} is formed by two
rapidly rotating em fields resulting in a desired time-averaged
phase profile.

We assume that $\sin(2kx)$ and
$\varphi(x)$ in Eq.~{(\ref{ome})}  are approximately constant over the
spatial overlap area of neighboring lattice site atom wavefunctions. Then
the sine function is approximately $\pm 1$  at each overlap area:
\beq
\kappa\simeq i\eta[(-1)^n+\varphi(x)]\,. \label{ome2}
\eeq
For notational simplicity, we take
$\kappa^*=-\kappa$, and $\eta$ real.

In this paper we study the Hamiltonian (\ref{ham}) [with
$\Omega(\rv)$ defined in Eqs.~{(\ref{ome})} and~(\ref{ome2})] only in the
continuum limit, where it can be transformed to the relativistic Dirac
Hamiltonian (\ref{dir}) exhibiting fractional charge
\cite{JAC76}. It may be shown that the fractional eigenvalues also emerge
in a more general case, but the continuum field theory is amenable to
a simpler description. The continuum limit corresponds to the
linearization of the fermionic band structure and becomes accurate in the
dilute gas limit, where the atomic correlation length is much larger than
the lattice spacing. In the continuum limit we write the fermionic
annihilation operators as continuous functions of the lattice spacing
$d\equiv \lambda/4$:
\beq
\alpha_j\equiv\sqrt{2d}\,u(j d),\quad \beta_j\equiv\sqrt{2d}\,v(j d)\,.
\eeq
Then the continuum limit proceeds exactly as in the polymer case
\cite{NIE86}. To
leading order in small $d$ we obtain
\bea H/\hbar &=& 2i d^2 \eta \sum_n [u^\dagger(n d)v'(n d) + v^\dagger(n
d)u'(n d)]\nonumber\\ &&+ {\delta d\over2}\sum_n [u^\dagger(n d) u(n
d)-v^\dagger(n d) v(n d)]\nonumber\\ &&+2i\eta d \sum_n\varphi(n d)
[u^\dagger (n d) v(n d)+ v^\dagger (n d) u(n d)]\,.
\label{hamcon}
\eea
Here $u'(n d)$ denotes a discrete spatial derivative of $u$.
In the continuum limit we replace $n d\rightarrow x$ and
$d\sum_n\rightarrow\int dx$. By introducing the spinor $\Psi(x)\equiv
[u(x) \,v(x)]^T$ and the transformation: $\Psi\rightarrow
\exp{(i\pi\sigma^3/4)}\Psi$, we may express Eq.~(\ref{hamcon}) as the
relativistic Dirac Hamiltonian (\ref{dir}), when we identify
$c=\lambda\eta/2$, $g=2\eta$, and $m=2\hbar\delta/(\lambda^2\eta^2)$. In
this case the spinor components refer to the two internal atomic levels.
Note that for nonzero detuning $\delta$ the system is not
charge conjugation symmetric, and the eigenvalues can have any
fractional value. The ratio between the coupling strength $\gamma\eta$
and $\delta$ therefore  determines the fractional part of the particle
number. In experiments this  could be engineered accurately, allowing a
controlled way of preparing  the fractional part of the eigenvalues.

The crucial part of our proposal for fractional particle number
is the em field $\varphi(x)$ in Eq.~{(\ref{ome})}. This is very different
from the fermion particle number fractionization in polymers, as our
fermionic fields are not coupled to a bosonic matter field
with a domain-wall soliton. Instead, the coherent em field with a
topologically appropriate phase profile is coupled to the FD atoms via
internal transitions. This results in the quantization of the FD atomic
gas with nontrivial topological quantum numbers corresponding to the
soliton sector of the relativistic 1+1 quantum field theory models of
fractionization. On the other hand, a spatially constant phase profile
$\varphi(x)$  represents the FD vacuum sector exhibiting integer particle
numbers and no bound state.

The normalizable bound state, which plays an important role in
the fractionization, belongs to only one of the two fermionic
components, independently of the shape or the position of the phase
kink: for the solitonlike phase kink,
$\varphi(h)=-\varphi(-h)=\gamma$, for all $h\gg d$, only level $\up$
is occupied, and for the antisolitonlike phase
kink, $\varphi(h)=-\varphi(-h)=-\gamma$, only level
$\down$ is occupied \cite{NIE86,HEE88}.
When $m=0$, the bound state has zero energy. Its spatial profile
$\sim\exp{(-\gamma |x|/d)}$ (for a sharp kink at $x=0$) depends on
the relative strength of the superposed em fields, according to
Eq.~{(\ref{ome})}, determining $\gamma$. Unlike in the polymer case,
where the size of the bound state is fixed, in the atomic case this
could be varied experimentally.

Because the local density of states is conserved, a zero-energy mode
creates a fractional deficit of states in both the valence and the
conduction bands. In the presence of charge conjugation symmetry, the
density of states is a symmetric function of the energy, and  both
bands have a decifit of one-half a state. By assigning the atoms in the
conduction and valence bands as ``particles" and ``antiparticles",
respectively, we can interpret the fractional particle number operator as
the occupation number difference between the bands. However, because the
zero-mode always occupies only one spin component at a time, also the
spin is fractionized. As an example, the two species may correspond to
the eigenstates of a single-particle spin operator along the $z$ axis,
$\sigma_z$, with eigenvalues $\pm1$. Then the eigenvalues of the
many-particle operator $S_z\equiv \sum_n \sigma_z^{(n)}$, localized around
the soliton, are fractional with vanishing fluctuations.

The total number of atoms, of course, must remain an integer and any
realistic optical lattice has a finite size. For every fractional
particle number located at the phase kink (forming a soliton), some
fractional charge is distributed at the boundary of the atomic cloud, or
is associated with an antikink. Although the fluctuations of the
fractional eigenvalues do not then in a finite lattice exactly vanish,
it can be shown that the fluctuations decay exponentially as a
function of the size of the system and can be considered negligible,
if the size is much larger than the atomic correlation length
\cite{kivelson,HEE88}. Therefore, {\it every} localized measurement of
the particle number around the phase kink can yield a fractional
result.

In experiments on fractional fermion number we may detect the
bound state or measure a fractional expectation value and
determine its fluctuations to ascertain that they are compatible
with fractionization. The FD gas exhibits a gap $\sim 2\eta$ in
the excitation spectrum. The phase kink creates a bound mode at
the center of the gap, hence excitations at half the gap energy \cite{zero}.
These midgap transitions~\cite{HEE88} could be probed in resonance
spectroscopy. The bound state also alters the dynamical structure
factor, which may be observed via light scattering~\cite{JAV95}.
The optical signal may be magnified by simultaneously preparing
many phase kinks. The fractional particle number could be detected
by measuring the occupation numbers of the individual lattice
sites. For instance, a magnetic field gradient may be introduced
that causes a detectable change in the spin flip frequency from
site to site~\cite{kas}. Finally, an off-resonance optical probe
couples to atom density~\cite{JAV95}. Fluctuations in the
scattered light should therefore convey information about not only
the particle number, but also about its fluctuations.

We can also generalize the proposed scheme to particle number
fractionization in higher dimensional models in 2D or 3D optical
lattices with atoms coupled to em fields exhibiting phase profiles
similar to topological defects or textures \cite{RUO01}.
In relativistic 2+1 D
quantum field theory, a fermionic field coupled to a bosonic field
exhibiting a vortex profile results in fermion particle number
fractionization \cite{NIE86}. In our approach this would correspond to a
2D optical lattice of FD atoms coupled by an em field with a nonvanishing
phase winding around any closed loop circulating the axis of a vanishing
field amplitude. This is also similar to the Raman field used in the
JILA experiments to couple two atomic condensates in order to create
vortices \cite{MAT99}. We may find an analogy to relativistic
3+1 D quantum field theory in the presence of a nonvanishing SU(2)
topological charge of the bosonic field by means of the em field configurations proposed in
Ref.~\cite{RUO01} to engineer Skyrmion textures in Bose-Einstein
condensates. Analogous techniques could possibly also be used to
couple FD atoms to em fields with the structure of 3D monopole defects
\cite{STO01}.

We have considered the em background field as a coherent classical field.
This is consistent with the adiabatic approximation in the fermionic
particle number fractionization, in which case the quantum fluctuations
of the bosonic soliton are ignored \cite{NIE86}. For a two-photon optical
Raman coupling between the two species, the decoherence rate per atom is
determined by the Rayleigh scattering rate $\sim\Gamma \max(|{\cal
R}_1|^2, |{\cal R}_2|^2)/\Delta^2$, where $\Gamma$ denotes the natural
linewidth. This can be reduced by increasing the detuning
$\Delta$, provided that at the same time a sufficient laser intensity for
the required tunneling rate $\Omega\sim{\cal R}_1{\cal R}_2/\Delta$ is
available. In the case of a rf coupling the quantum effects of the em
field  can be safely ignored on the time scale of the experiments.

This work was financially supported by the EPSRC, the U.S. DOE., the U.S.
NSF, and NASA. We acknowledge discussions with M. Kasevich and C.M. Savage.


\begin{references}

\bibitem{jackiw} R. Jackiw, Dirac Prize Lecture, hep-th/9903255.

\bibitem{anderson} P. W. Anderson, Phys. Today {\bf 50}, 42 (October
1997).

\bibitem{JAC76} R. Jackiw and C. Rebbi, Phys. Rev. D {\bf 13}, 3398
(1976).

\bibitem{NIE86} For a review, see: A. Niemi and G.
Semenoff, Phys. Reports {\bf 135}, 99 (1986), and references therein.

\bibitem{OLS98} The particle number fractionization could also be
realized with tightly confined 1D bosonic atoms in the Tonks gas
regime, where the impenetrable bosons obey FD statistics, M. Olshanii, Phys. Rev. Lett. {\bf 81}, 938 (1998).


\bibitem{SU79} W.P. Su, J.R. Schrieffer, and A.J. Heeger, Phys. Rev.
Lett. {\bf 42}, 1698 (1979).

\bibitem{HEE88} A.J. Heeger {\it et al.}, Rev. Mod. Phys. {\bf 60}, 781
(1988).

\bibitem{laughlin} R. B. Laughlin, H. St\"ormer and D. Tsui, Rev. Mod.
Phys. {\bf 71}, 863 (1999).

\bibitem{AND01} B.P. Anderson {\it et al.}, Phys. Rev. Lett. {\bf 86},
2926 (2001).


\bibitem{MAT99} M.R. Matthews {\it et al.}, Phys. Rev. Lett. {\bf 83},
3358 (1999).


\bibitem{DEM99} B. DeMarco and D.S. Jin, Science {\bf 285}, 1703 (1999);
A.G. Truscott {\it et al.}, Science {\bf 291}, 2570 (2001); F. Schreck
{\it et al.}, Phys. Rev. Lett {\bf 87}, 080403 (2001).


\bibitem{AND98} B.P. Anderson and M.A. Kasevich, Science {\bf 281}, 1686

(1998); C. Orzel {\it et al.}, Science {\bf 291}, 2386 (2001); M. Greiner
{\it et al.}, Phys. Rev. Lett. {\bf 87}, 160405 (2001).


\bibitem{JAC98} D. Jaksch {\it et al.}, Phys. Rev. Lett. {\bf 81}, 3108
(1998); J. Javanainen, Phys. Rev. A {\bf 60}, 4902 (1999).





\bibitem{RUO01} J. Ruostekoski and J.R. Anglin, Phys. Rev. Lett. {\bf
86}, 3934 (2001).


\bibitem{GW} J. Goldstone and F. Wilczek, Phys. Rev. Lett. {\bf 47},
986 (1981).



\bibitem{opt} With simple modifications we could also use, e.g., the
optical phase profile $\Omega(\rv)= i{\cal
V}(\rv)\,[1+\sin(2kx)\varphi(x)]$.


\bibitem{JAV95} J. Javanainen and J. Ruostekoski, Phys. Rev. A {\bf 52},
3033 (1995).

\bibitem{PU01} H. Pu {\it et al.}, Phys. Rev. A {\bf 63}, 063603 (2001).


\bibitem{kivelson} S. Kivelson and J. R. Schrieffer, Phys. Rev. B {\bf
25}, 6447 (1982); R. Jackiw {\it et al}, Nucl. Phys. B{\bf 225}, 233
(1983).

\bibitem{zero} In a finite-size lattice the midgap mode may also exist in the vacuum state. However, this is always spatially separated from the soliton bound state.

\bibitem{kas} This was pointed out by M.A. Kasevich.

\bibitem{STO01} H.T.C. Stoof {\it et al.}, Phys. Rev.
Lett. {\bf 87}, 120407 (2001).



\end{references}
\end{document}